\documentclass{article}

\usepackage{arxiv}

\usepackage[utf8]{inputenc} 
\usepackage[T1]{fontenc}    
\usepackage{hyperref}       
\usepackage{url}            
\usepackage{booktabs}       
\usepackage{amsfonts}       
\usepackage{nicefrac}       
\usepackage{microtype}      
\usepackage{lipsum}
\usepackage{graphicx}

\usepackage[utf8]{inputenc}
\usepackage{enumerate}
\usepackage{graphicx}
\usepackage{inputenc}
\usepackage[linesnumbered,ruled,vlined]{algorithm2e}
\usepackage{amsmath}
\usepackage{algorithmic}
\usepackage{upquote}
\usepackage{xcolor}
\usepackage{comment}
\usepackage{subfigure}
\usepackage{mathtools}
\usepackage{multirow}
\usepackage{lscape}
\usepackage{fancyhdr}
\graphicspath{ {./images/} }
\newcommand\blfootnote[1]{%
  \begingroup
  \renewcommand\thefootnote{}\footnote{#1}%
  \addtocounter{footnote}{-1}%
  \endgroup
}

\title{An Approximate Carry Estimating Simultaneous Adder with Rectification}

\author{
 Rajat Bhattacharjya* \\
Indian Institute of Information Technology Guwahati \\
Assam, India. \\
  \texttt{rajat.iiitg@gmail.com} \\
   \And
Vishesh Mishra* \\
Indian Institute of Information Technology Guwahati \\
Assam, India. \\
  \texttt{visheshmishra432@gmail.com} \\
  \And
 Saurabh Singh \\
Indian Institute of Information Technology Guwahati \\
Assam, India. \\
  \texttt{saurabh.s99100@gmail.com} \\
   \And
Kaustav Goswami \\
Indian Institute of Information Technology Guwahati \\
Assam, India. \\
  \texttt{kaustavgoswami.2013@gmail.com} \\
  \And
 Dip Sankar Banerjee \\
Indian Institute of Information Technology Guwahati \\
Assam, India. \\
  \texttt{dipsankarb@iiitg.ac.in} \\
}

\begin{document}
\maketitle
\begin{abstract}
Approximate computing has in recent times found significant applications towards lowering power, area, and time requirements for arithmetic operations. Several works done in recent years have furthered approximate computing along these directions. In this work, we propose a new approximate adder that employs a carry prediction method. This allows parallel propagation of the carry allowing faster calculations. In addition to the basic adder design, we also propose a rectification logic which would enable higher accuracy for larger computations. Experimental results show that our adder produces results 91.2\% faster than the conventional ripple-carry adder. In terms of accuracy, the addition of rectification logic to the basic design produces results that are more accurate than state-of-the-art adders like SARA~\cite{sara} and BCSA~\cite{bcsa} by 74\%.
\end{abstract}


\blfootnote{$^{*}$ Both authors contributed equally to this research.}
\blfootnote{This work has been accepted at 30th ACM Great Lakes Symposium on VLSI (GLSVLSI) 2020 as a regular paper.}
\section{Introduction}

Approximate computing in recent years has garnered significant attention due to the massive data deluge and requirements for fault-tolerant real-time computing. Towards this, there have been several techniques that have been proposed which have shown that often an approximate result with provable error bounds are desirable rather than computing correct result from scratch which can take significantly higher time. With massive advancements in semiconductor technologies in recent years, digital circuits have also become more vulnerable to variations. These variations have made accurate results more difficult to be ensured~\cite{approx}. 

Techniques employed for analyzing large data and employing machine learning methods often rely on approximations to quickly model the available data. For the design of semiconductor devices, such methods are often feasible in the sense that if the applications are in itself employing some level of approximation, then the results provided by hardware can also assist in approximation. This can potentially lead to much faster computations, simpler hardware, and additional power benefits. 

Works done towards the implementation of approximations on hardware~\cite{sara, ansari} in recent years have proposed techniques that have shown that approximation shows acceptable results on the end applications. An approximation can be done on several levels and works focusing on both full systems approximation~\cite{sys1} and software level approximations~\cite{soft2} have been shown. Arithmetic circuits such as adders and multipliers form the basic building blocks in all such approximate systems. On hardware, techniques proposed towards achieving approximate results are mostly non-configurable which means that it is not possible to trade-off the approximated errors with the amount of extra logic that needs to be implemented.
In this work, we propose a new solution for approximate additions via a simultaneous carry estimation technique (CESA). Additionally, we fine-tune our design to ensure that only a marginal amount of extra space is consumed on the die. Also, we propose a propagating error rectification logic (PERL) which would yield higher accuracy compared to the basic adder design (CESA) for larger computations. We name this modified design as CESA-PERL (Carry Estimating simultaneous Adder with Propagating Error Rectification Logic). An intuitive look at how adders work in hardware (like the popular Ripple-Carry adder) shows that the opportunity for any approximate result depends on how early a possible point of error could be detected. These are the potential points of good approximation as these carries that are generated after every pair of bits are added. So, a naive way to take care of the minimal error accumulation at the carries could be via sequential processing of the carries as and when they are generated. Although on close observation, we can understand that once the instruction for addition is invoked, the register gets results after a finite amount of delay. So, if we  look at all the carries generated in parallel and ensure that the error is not accumulated beyond a certain threshold, then the result can be quickly generated and accurately estimated.


  \begin{figure*}[tb]
        \begin{center}
                \includegraphics[width=0.98\textwidth]{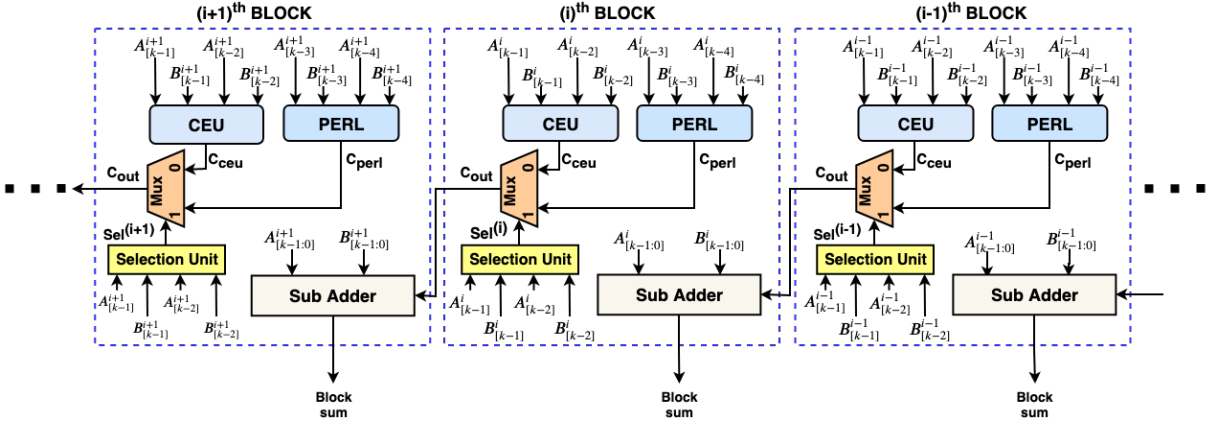}
                
                \caption{Cascaded blocks of CESA-PERL}
                \label{fig:carry-gen}
        \end{center}
  \end{figure*}


Our main motivation behind this work is achieving faster computations through approximate additions which incur a lower area overhead. Since accuracy is a function of the chip area, we exploit the fundamental intuition that errors cumulatively build across parallel addition blocks. If we can track this propagation, we can correct them trivially which will lead to much better accuracy. This is what we achieve through the design of our naive approximate adder called CESA which is further optimized through the rectification of the propagating errors through the CESA-PERL design (discussed in Section~\ref{sec:method}). We show that on the application's end, we achieve speedup of around 2.83x over accurate adders, at an additional space requirement of 12.5\%. When compared to other state-of-the-art approximate adders, we observe that our design outperforms existing designs that consume similar area overheads by 74\% on an average (discussed in Section~\ref{sec:eval}).

\section{Developmental Ideas}
\label{sec:Preliminaries}
      
      
In this section, we discuss the preliminary ideas behind our designs. The basic architecture of the proposed adder depicted in Figure~\ref{fig:carry-gen} comprises of $[n/k]$ k-bit summation blocks which generate the partial result in a non-blocking parallel manner. If two n-bit input operands are taken for approximate additions, then the proposed adder CESA-PERL will divide the input operands into $n/k$ equal segments, called sub-inputs. The sub-inputs are then fed to a k-bit sub-adder which is part of the k-bit summation block. The k-bit summation block also contains the CEU (Carry Estimate Unit), PERL (Propagating Error Rectification Logic), and an SU (Selection Unit).    

In Figure~\ref{fig:carry-gen}, the carry input of $i^{th}$ sub-adder is selected through $(i-1)^{th}$ SU which selects the output of CEU if it estimates the carry input correctly. Also, the output of PERL in case CEU generates an error in carry estimation. Since the carry input of the $i^{th}$ sub-adder is not generated through the actual carry chain mechanism but selected directly by SU, the results are approximate in nature. In our proposed adder, the estimated carry input of  $i^{th}$ sub-adder is given by,
    \begin{equation}
    C_{out} =\overline{Sel^{(i-1)}}.C_{ceu} + Sel^{(i-1)}.C_{perl}
    \label{eqn:EQ1}
    \end{equation}
In equation \ref{eqn:EQ1}, $C_{out}$ beside being the carry input of the $i^{th}$ sub adder, it is also the carry output of the $(i-1)^{th}$ sub adder. Here,  $\overline{Sel^{(i-1)}}.C_{ceu} $ is the output signal of the select (CEU) unit and $Sel^{(i-1)}.C_{perl}$  is the output signal of the select unit (PERL). Also, $Sel^{(i-1)}$ is the $(i-1)^{th}$ Selection Unit (SU) whose logic is given by, 
   \begin{equation}
   \label{eqn:EQ2}
   Sel^{i-1}=(A^{i-1}_{k-1}\oplus B^{i-1}_{k-1}).(A^{i-1}_{k-2}\oplus B^{i-1}_{k-2})
   \end{equation}
   subsequently $C_{ceu}$, $C_{perl}$ are respectively given by,
    \begin{equation}
    \label{eqn:EQ3}
    C_{ceu} = A^{i-1}_{k-1}.B^{i-1}_{k-1} + A^{i-1}_{k-2}.B^{i-1}_{k-2}( A^{i-1}_{k-1}+ B^{i-1}_{k-1} )
    \end{equation}
    \begin{equation}
    \label{eqn:EQ4}
    C_{perl} = A^{i-1}_{k-3}.B^{i-1}_{k-3} + A^{i-1}_{k-4}.B^{i-1}_{k-4}( A^{i-1}_{k-3}+ B^{i-1}_{k-3} )
    \end{equation}
    Here in equations \ref{eqn:EQ2}, \ref{eqn:EQ3} and \ref{eqn:EQ4}, $A^{i-1}_{k-1}$ and $B^{i-1}_{k-1}$ can be interpreted as $(k-1)^{th}$ input bits of $(i-1)^{th}$ sub-adder. The other terms used in these equations can also be interpreted in a similar manner. 
    
\subsection{CESA}
Carry Estimating Simultaneous Adder (CESA) contains various sub adders which compute individual block sum in parallel. The Carry Estimate Unit (CEU) generates the carry input for the summation block based on the two most significant bits of the summation block previous to it. Since four input bits are involved namely $A^{i-1}_{k-1}$, $B^{i-1}_{k-1}$, $A^{i-1}_{k-2}$ and $B^{i-1}_{k-2}$, so total $2^4=16$ permutations arise. The idea behind carry estimation by CEU is as follows: 
        \begin{itemize}

        \item If both $A^{i-1}_{k-1}$ and $B^{i-1}_{k-1}$ are $(0,0)$ respectively, then the carry input for the $i^{th}$ block would also be $0$ irrespective of previous carry-ins. This will account for four binary combinations. Since $A^{i-1}_{k-2}$ and $B^{i-1}_{k-2}$ can still vary over 0 and 1. 
        
         \item If both $A^{i-1}_{k-1}$ and $B^{i-1}_{k-1}$ are $(1,1)$ then the carry input for the $i^{th}$ block would also be $1$ irrespective of previous carry-ins. This also includes four binary combinations.
         
        \item If $A^{i-1}_{k-1}$ and $B^{i-1}_{k-1}$ are (1,0) or (0,1), then the carry input for the $i^{th}$ block would depend on the previous bits. If the previous bits $A^{i-1}_{k-2}$ and $B^{i-1}_{k-2}$ are $(0,0)$, then carry input would be $0$ and if previous bits $A^{i-1}_{k-2}$ and $B^{i-1}_{k-2}$ are $(1,1)$ then it would be $1$. 
        
        \item For the remaining cases, to accurately determine carry input, a further backward input bit traversal would be required. So in all four remaining cases, carry input is approximated as $0$ without any backward traversal. This approximation leads to tolerable error as the actual carry input may not be $0$.
        
\end{itemize}

Let $C_{radd}$ be the accurate carry input for $i^{th}$ sub-adder which would have been generated through the carry chain mechanism in a traditional ripple carry adder, then based on the above-mentioned cases, the carry input is always accurate in $12$ out of $16$ cases. So, the probability of correct carry bit estimation is,
    \begin{equation}
     P(C_{ceu}=C_{radd}) = 12/16=3/4 
     \label{eqn:EQ5}
    \end{equation}
Clearly, the error occurs only when carry estimation is false. So by making the use of equation \ref{eqn:EQ5}, the probability that the result generated by CESA is always error-prone is:
        \begin{equation}
        \label{eqn:EQ6}
        P(C_{ceu}\neq C_{radd})=1 - P(C_{ceu}=C_{radd})= 1/4
         \end{equation}
\subsection{CESA-PERL}
Now in equation \ref{eqn:EQ6}, the probability of getting error-prone results that comes out to be 1/4 can be interpreted as to be the worst case error rate of $25$\%,provided the number of summation blocks are minimum.

Since the worst case error rate is quite high, so in order to rectify the error, we propose an auxiliary rectification unit known as PERL.
When both input bits, $A^{i-1}_{k-1}$ and $B^{i-1}_{k-1}$ are (1,0) or (0,1) respectively and the CEU is unable to determine the carry input accurately, then instead of selecting $0$ as the approximate carry input, the output signal of PERL is selected which significantly lowers error-rates. 

Now the error will only occur if both units, namely CEU and PERL simultaneously fail in accurate determination of carry input. So the probability that the result generated by the adder is always error-prone is:
        \begin{equation}
        \label{eqn:EQ7}
        P(C_{out}\neq C_{radd})=P(C_{ceu}\neq C_{radd})\times P(C_{perl}\neq C_{radd})= 1/16
        \end{equation}
It is clear from equation \ref{eqn:EQ7} that the chances of getting error-prone results decrease significantly after the addition of PERL. Although adding PERL introduces some additional area overheads, the benefits that it provides towards accuracy clearly outweigh the nominal area overheads.
\section{Design and Implementation}
\label{sec:method}

In this section, we briefly discuss the solution mechanism along with the hardware level designs that are proposed.
                       \begin{figure*}[tb]
        \begin{center}
            \mbox{ 
            \hspace{-1.0ex}
            \subfigure[Error Rate (\%)]
                {
                \label{fig:er}
                \includegraphics[width=0.32\textwidth]{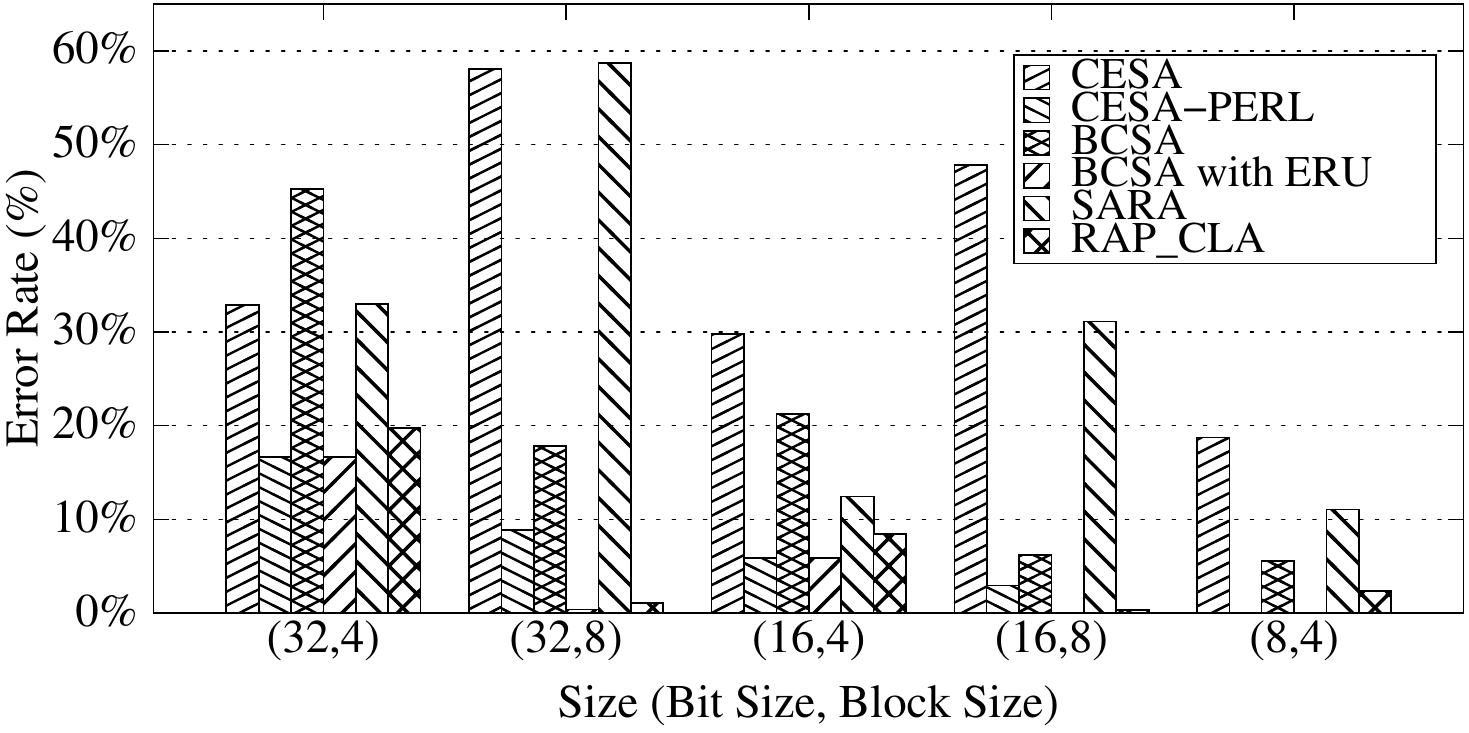}
                }
            \hspace{-2.0ex}
            \subfigure[Mean Relative Error Distance (MRED)]
                {
                \label{fig:mred}
                \includegraphics[width=0.32\textwidth]{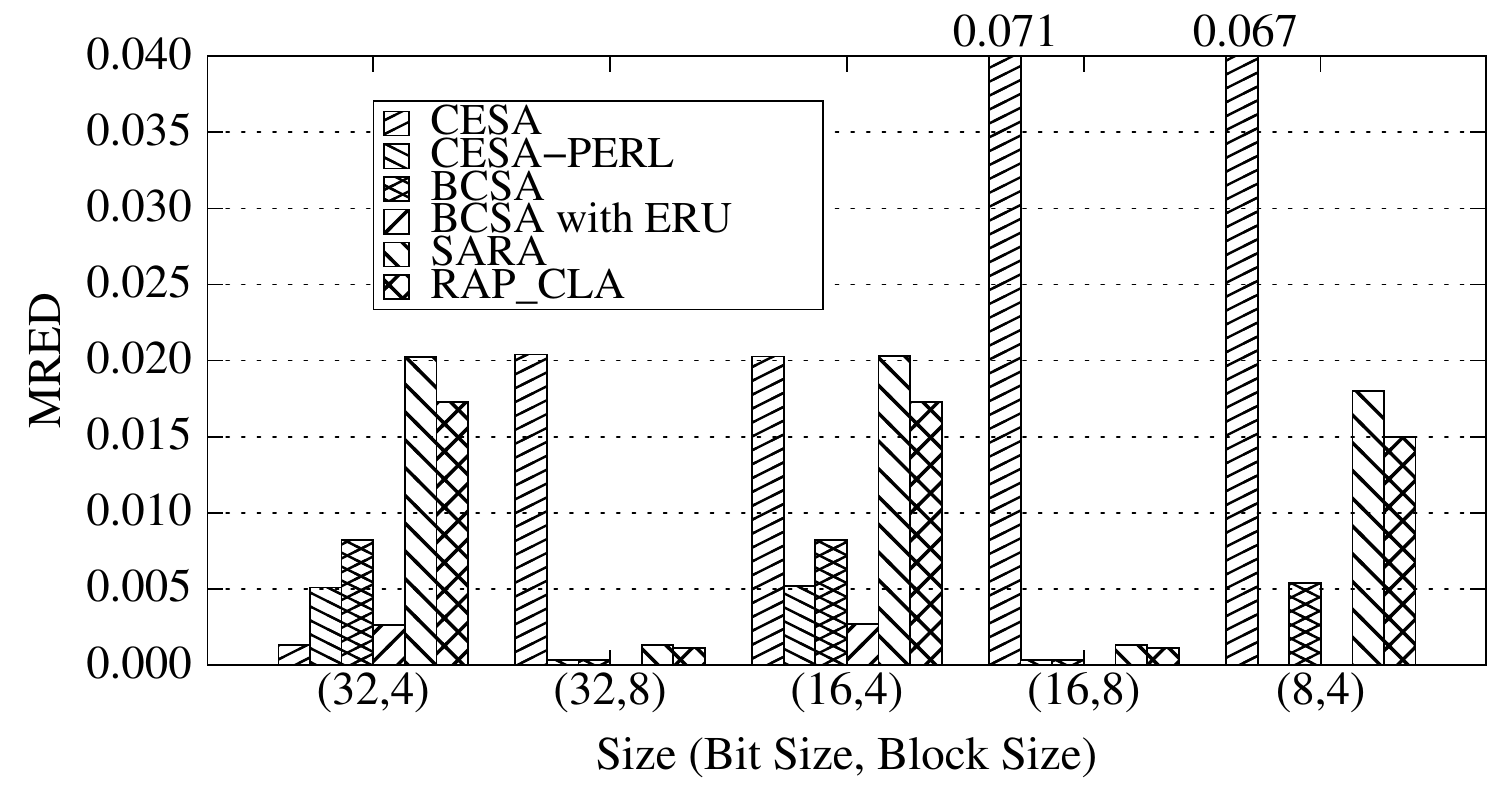}
                }
            \hspace{-2.0ex}
            \subfigure[Mean Error Distance (MED)]
                {
                \label{fig:med}
                \includegraphics[width=0.32\textwidth]{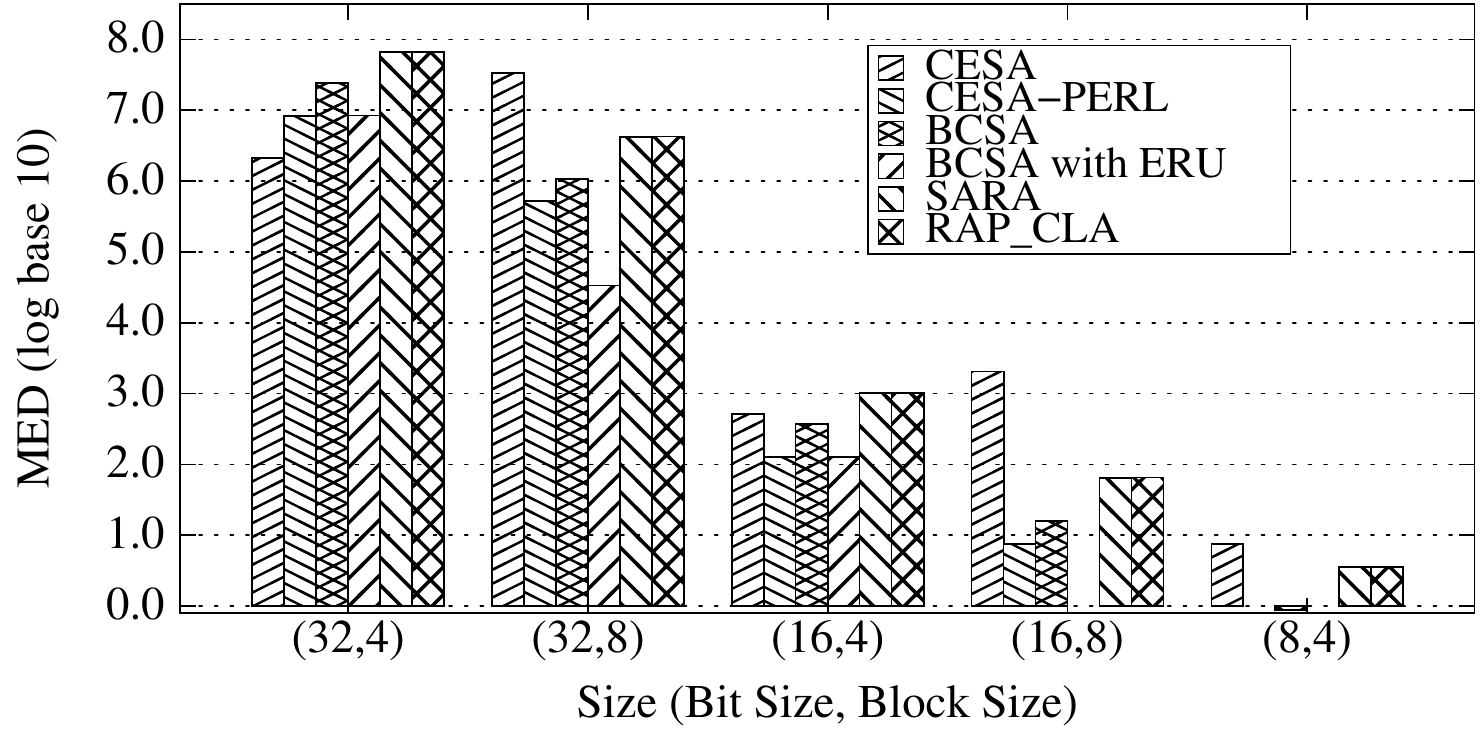}
                }
            }
            \caption{Error Metrics}
            \label{fig:error-metrics} 
        \end{center}
    \end{figure*}         
\subsection{Hardware Design}
\label{sec:hardware}

Our proposed adder is designed with the sole motivation of providing higher accuracy at the cost of lower on-chip area and minimal power consumption. The basic design is based on the divide and conquer technique in which we divide the $n$ bit input operand into $n/k$ equal segments called sub inputs. The sub inputs are then fed to sub adders which are the part of summation blocks. The summation blocks of size $k$ can be evaluated in a parallel manner to calculate their individual sum. Since all summation blocks are independent of each other, they will produce results at the same instant after a certain delay. Meanwhile, just before the beginning of computations, the carry input of these blocks is selected through SU. As discussed in section \ref{sec:Preliminaries}, SU selects the output of either CEU or PERL based on various cases. CEU, PERL and SU generate results in a non-blocking manner so that it can be done concurrently with the sum calculation. This eliminates the added latency in the generation of carry input, hence producing the result in much lesser time.
    \begin{algorithm}
		\caption{ SummationBlock( $A^{i-1}_{[k-1:0]}$, $B^{i-1}_{[k-1:0]}$, $C{in}$ )}
		\label{alg:block}
	\begin{algorithmic}[1]
		\STATE Initialize: sum[n]
		\STATE $C_{ceu}$ = CEU( $A^{i-1}_{[k-1]}$,$B^{i-1}_{[k-1]}$,  $A^{i-1}_{[k-2]}$,   $B^{i-1}_{[k-2]}$ )
		\STATE $C_{perl}$ = PERL( $A^{i-1}_{[k-3]}$,$B^{i-1}_{[k-3]}$,  $A^{i-1}_{[k-4]}$,   $B^{i-1}_{[k-4]}$ )
		\STATE $Sel$ = SU( $A^{i-1}_{[k-1]}$, $B^{i-1}_{[k-1]}$, $A^{i-1}_{[k-2]}$, $B^{i-1}_{[k-2]}$ )
		\STATE $C_{out} =\overline{Sel}.C_{ceu} + Sel.C_{perl}$
		\FOR{j $\leftarrow$ 0 to k in parallel }
        \STATE sum[j] = $A^{i-1}_{[j]}$ $\oplus$ $ B^{i-1}_{[j]}$ $\oplus$ $ C_{in}$
        \STATE $C_{in}$ = ( $A^{i-1}_{[j]}$ $\cdot$ $B^{i-1}_{[j]}$) + (($A^{i-1}_{[j]}$ $\oplus$ $ B^{i-1}_{[j]}$) $\cdot$ $ C_{in}$ )
		\ENDFOR
		\STATE return (sum, C$_{out}$)

	\end{algorithmic}
		\end{algorithm}

                
                

Algorithm~\ref{alg:block} shows how estimation of carry input for $i^{th}$ block and sum calculation in $(i-1)^{th}$ summation block takes place simultaneously. Here $Sel$,  $C_{ceu}$ and $C_{perl}$ are computed using equations \ref{eqn:EQ2}, \ref{eqn:EQ3} and \ref{eqn:EQ4} respectively. Each summation block returns sum bits and a carry-out $( C_{out} )$. This $C_{out}$ acts as a carry input for the next block. The carry input for the first block is initialized as zero while the other summation blocks take the carry-out of the previous block as their carry input. In this way, all the blocks are evaluated in a parallel manner and each block follows Algorithm~\ref{alg:block} for generating results. Now we will briefly discuss various components of the summation block. 


\subsubsection{ \textbf{Carry Estimate Unit (CEU)}} 

This unit estimates the carry input of the next bock based on the two most significant bits of the previous block. It produces the output after two gate-level delays which are faster than the delay provided by a single full adder. Equation~\ref{eqn:EQ3} describes the logic behind it. 

\subsubsection{ \textbf{Propagating Error Rectification Logic (PERL)}}

The hardware design of PERL is exactly the same as that of CEU but a different set of input are fed to PERL. In case CEU wrongly estimates the carry input using two most significant bits of the block then the other two most significant bits adjacent to previous ones are fed to PERL. As discussed in equation \ref{eqn:EQ7}, the chances of estimating the carry input incorrectly decreases after the inclusion of PERL in hardware design. In this way, PERL rectifies the propagating error due to false carry estimation.
\subsubsection{ \textbf{Selection Unit (SU)}}
\label{sec:su}
The SU selects the output signal of CEU in case it estimates the carry input correctly and the output signal of PERL if the carry estimated through CEU was false and required rectification. Equation \ref{eqn:EQ2} showcases the logic of SU which is obtained through boolean simplification. It also generates the result after two gate-level delays ensuring in a faster selection of available carry input. Figure~\ref{fig:carry-gen} shows the three cascaded summation blocks of our proposed adder design. The logic for all individual units of summation block are generated through Boolean simplifications and thus the circuit for each unit is made as per logic expressions.

      
In our proposed adder CESA-PERL, the minimum block size is $4$ because a minimum of at least $4$ input bits are required to estimate the carry bit with error rectification via PERL. However, if the propagating error is ignored and PERL is not included in hardware design (as in case of CESA) then the minimum block size can be reduced to $2$ as carry estimate unit requires only $2$ input bits of the summation block in order to generate the carry input for next block.

Figures~\ref{fig:er}, \ref{fig:mred} and \ref{fig:med}  depict the error results. CESA-PERL shows the least error rate when the block size is lowest. Since the minimum block size can be four, the least error rate occurs when the input operands are divided into $n/4$ blocks. 

    \subsection{Compiler and ISA Extension}
     \label{subsec:compiler}
    
In order to facilitate the use of our proposed adder from software perspective, we propose two additional assembly level instructions, namely \textit{adx} and \textit{adxi} on existing ISAs. 
\textit{adx}, or \textit{approximately add}, approximately adds two numbers \textit{r1} and \textit{r2} explicitly on the CESA/CESA-PERL circuit. \textit{adxi}, or \textit{approximately add immediate} will function same as \textit{adx} but will include immediate as an operand. 
        

\section{Evaluation of the adder}
In this section, we evaluate CESA and CESA-PERL using standard metrics to measure error and the gains obtained in power and delay.
    \label{sec:eval}

\subsection{Accuracy Analysis}

CESA and CESA-PERL are compared to various other state-of-the-art approximate adders and all of them are evaluated based on standard error metrics. Error analysis was done using GNU Octave for $10^6$ random cases and averages taken over a dozen runs. The comprehensive results are shown in Figure~\ref{fig:error-metrics}. Error metrics such as ER(Error Rate)~\cite{metric}, MED (Mean Error Distance)~\cite{metric}, MRED (Mean Relative Error Distance)~\cite{metric} are used to compare our designs with SARA~\cite{sara}, RAP-CLA~\cite{rapcla}, BCSA and BCSA with ERU (Error Reduction Unit)~\cite{bcsa}.
        
For 8-bit numbers, CESA was found to be giving accurate results 85.94\% times whereas RAP-CLA was found to be giving accurate results 91.1\% times since RAP-CLA is extensively based on the carry-lookahead adder (shown in Figure~\ref{fig:er}). This is however at an area cost that is significantly higher than our solution as discussed in Section~\ref{sec:arrr}. SARA and BCSA give accurate results 82.25\% and 83.5\% respectively which is lesser than CESA but BCSA with ERU produces results 90.55\% times accurately. The good output of BCSA with ERU can be credited to the error reduction unit at the cost of an area overhead. For 16-bit numbers, CESA was found to be giving accurate results 70.1\% times whereas RAP-CLA gave accurate results 85\% times, SARA, BCSA, and BCSA with ERU gave accurate results 68.4\%, 70.6\% and 82.2\% times respectively. For 32-bit numbers too, CESA was found to be better than SARA and BCSA in terms of accuracy by 42.5\%.

For CESA-PERL, we find that error rates significantly drop as shown in Figure~\ref{fig:er}. The drop in error rates can be attributed to the use of PERL which is however at a minimal area overhead. On average we see that CESA-PERL is better than SARA and BCSA by 74\%.
        
\subsection{Hardware Evaluation}
        \begin{figure*}[tb]
        \begin{center}
            \mbox{ 
            \hspace{-1.0ex}
            \subfigure[Area ($\mu$m$^2$)]
                {
                \label{fig:area}
                \includegraphics[width=0.32\textwidth]{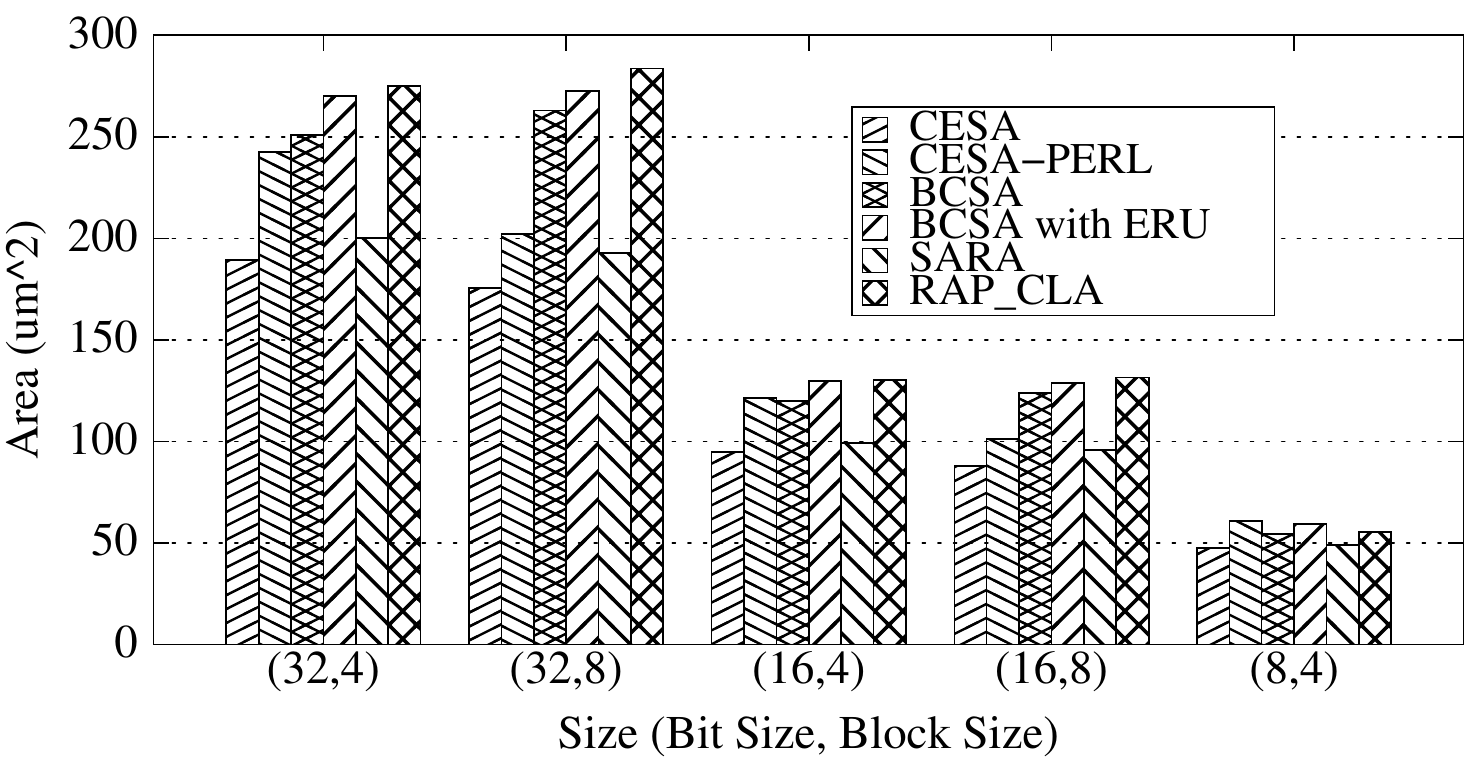}
                }
            \hspace{-2.0ex}
            \subfigure[Power ($\mu$W)]
                {
                \label{fig:power}
                \includegraphics[width=0.32\textwidth]{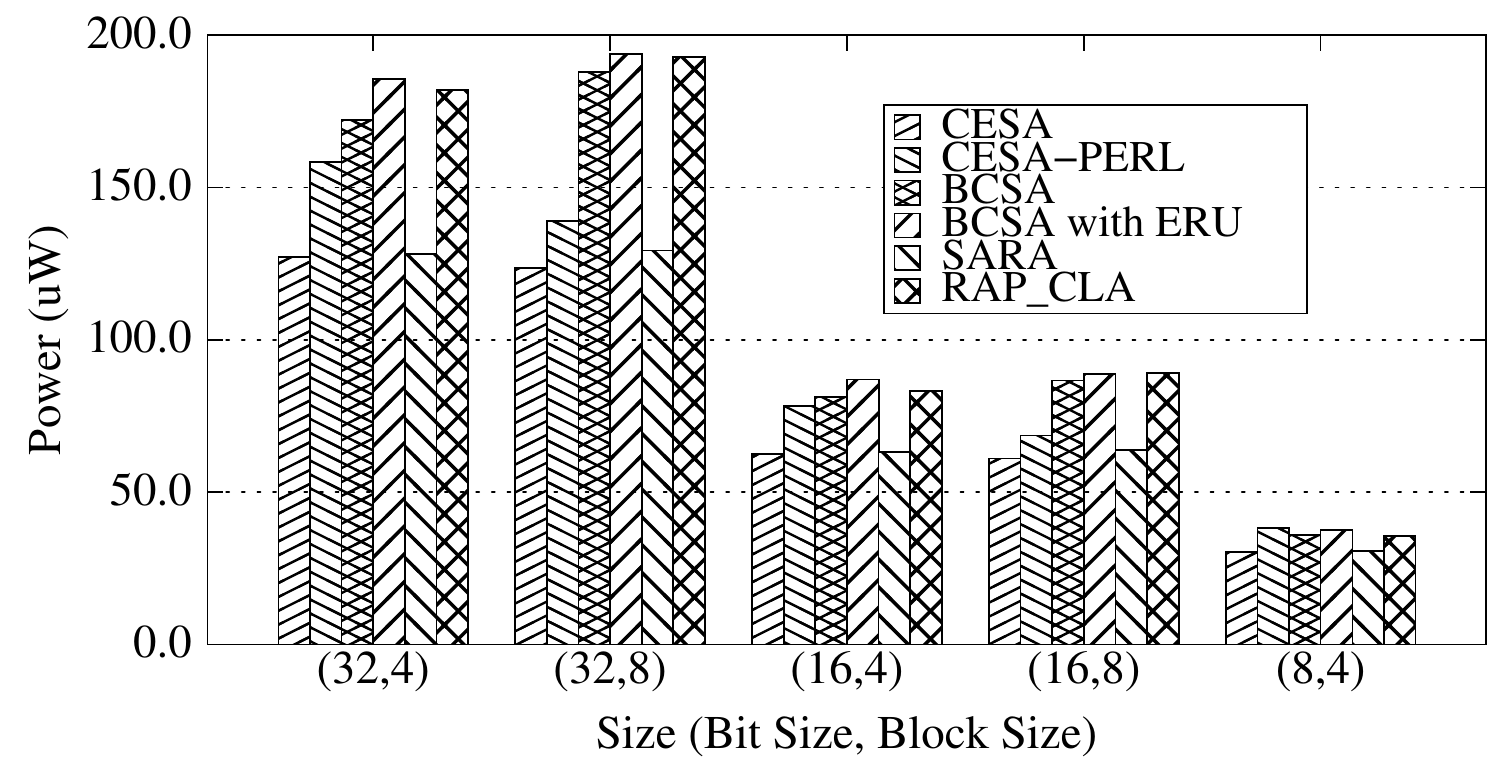}
                }
            \hspace{-2.0ex}
            \subfigure[Delay ($ns$)]
                {
                \label{fig:delay}
                \includegraphics[width=0.32\textwidth]{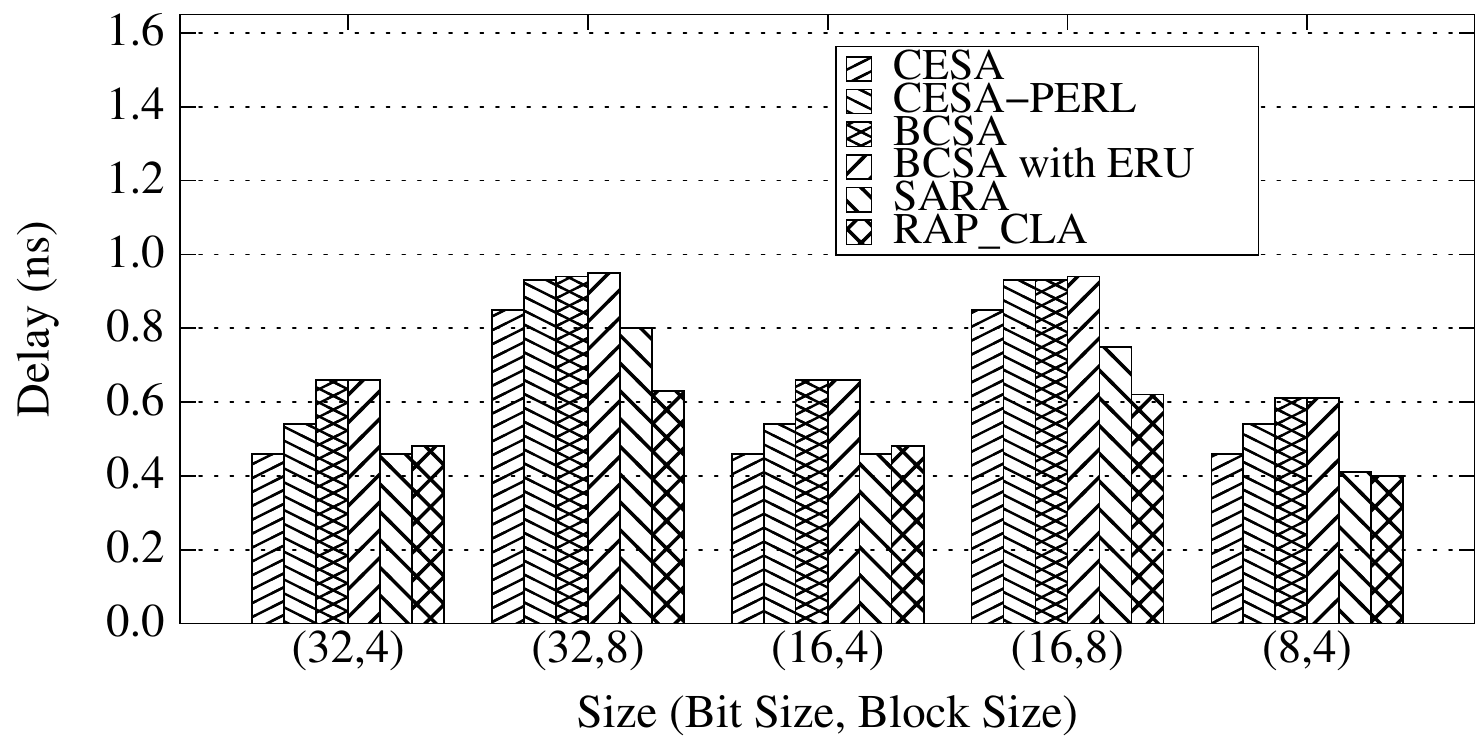}
                }
            }
            \caption{Hardware Evaluation}
            \label{fig:hw-parameters} 
        \end{center}
    \end{figure*}
We compare our adder CESA and CESA-PERL with SARA~\cite{sara}, RAP-CLA~\cite {rapcla}, BCSA and BCSA with ERU~\cite{bcsa} using Synopsys Design Compiler (DC) with The NanGate Open Cell Library (45nm technology node) on global operating voltage $1.1 $ V. All the adders were described using Verilog HDL. We add 8, 16 and 32-bit numbers for various block sizes thereby configuring accuracy. 
\subsubsection{\textbf{Delay:}}
For an 8-bit design, CESA shows faster output generation than SARA (by 2.99\%), BCSA (by 17.29\%) and BCSA with ERU (by 25.8\%). Similarly, for 16-bit and 32-bit designs, we find that CESA yields result faster as shown in Figure~\ref{fig:delay}. Higher delay of CESA compared to RAP-CLA can be attributed to the use of the carry-lookahead adder concept presented in it. On average we see that CESA's delay is 14.57\% lower than that of SARA, BCSA, and BCSA with ERU combined. When compared to the conventional ripple-carry adder, we find that CESA is 91.2\% faster than it when used in a best-case scenario. Due to the additional logic of rectification in CESA-PERL, we find that on average SARA and RAP-CLA outperform CESA-PERL by 26.4\%. Whereas when comparing with BCSA and BCSA with ERU, the delay statistics of CESA-PERL is better than them by 9.98\% across all configurations as shown in Figure~\ref{fig:delay}.
 
\subsubsection{\textbf{Area:}} 
\label{sec:arrr}
Investigating the statistics of area (shown in Figure~\ref{fig:area}), we see that for 8-bit additions CESA takes 10.51\% lesser area compared to SARA, RAP-CLA, BCSA and BCSA with ERU. More area of CESA compared to SARA is due to the use of additional circuitry for parallel carry estimation. Even though SARA also does a parallel estimation, our implementation looks at the vicinity of MSBs for carry estimation whereas SARA simply looks at the MSB. Coming to 16-bit architecture, we see that CESA takes less area than RAP-CLA (by 24.83\%), SARA (by 3.63\%), BCSA (by 20.64\%) and BCSA with ERU (by 27.43\%). Similarly, lesser area of CESA can be observed for the 32-bit design as well. On average, we see that the area overhead of SARA is actually more than CESA given the advantages that CESA gets when the carry estimation happens over longer distances. In CESA-PERL, the rectification logic (PERL) adds on to the area of the CESA. Even after the addition of PERL to CESA, area of CESA-PERL comes out to be 10.3\% lesser than that of RAP-CLA, BCSA and BCSA with ERU on an average.

\subsubsection{\textbf{Power:}}        
Power comparison in $\mu W$ is shown in Figure~\ref{fig:power}. In the case of 8-bit designs, we see that CESA takes 9.49\% less power than RAP-CLA, 1.90\% more power than SARA, 10.56\%, 19.44\% less power than BCSA and BCSA with ERU respectively. More power is consumed with respect to SARA in 8-bit design, given the parameter of larger area in CESA. For 16-bit additions, we find that CESA takes 17.33\% lesser power on an average compared to the other four adders. Finally, for 32-bit designs, we see that CESA takes 20.23\% less power on an average than RAP-CLA, SARA, BCSA, and BCSA with ERU combined. The lesser power that is consumed by CESA overall is due to the lesser amount of extra logic that is used by CESA for implementing the approximate adder in comparison to the others. We see that on average it takes 12.54\% less power in CESA-PERL than RAP-CLA, BCSA, and BCSA with ERU as shown in~Figure~\ref{fig:power}. Though CESA-PERL takes higher power than SARA, it is worth noting that the accuracy of SARA stands nowhere near to that of CESA-PERL.
\subsection{Systems Level Simulations}
\label{subsec:systems}

We have selected 7 integer SPEC CPU2006 benchmarks namely bzip2, sjeng, astar, libquantum, mcf, hmmer and omnetpp to evaluate speedup of our proposed approximate adder on end applications. We have used the delay values obtained using Synopsys Design Compiler for all 32 bit configurations. Using the same, we have modified the addition parameters in GEM5 to measure the runtime of end applications. This experiment is geared only towards the measurement of speedup obtained at the application end and not the program's correctness. Simulation was done for 1 billion instructions. We have simulated the benchmarks on GEM5~\cite{gem5} using a system with 4 out-of-order CPU cores with frequency 2 GHz each and a DRAM size of 4 GB. The system had three levels of cache (L1, L2, L3) of sizes 64 KB, 512 KB and 4 MB 
respectively. 


\section{Applications Evaluation}
In this section we briefly evaluate the performance of some applications that can potentially benefit from CESA and CESA-PERL.


\subsection{Image Processing: Gaussian smoothing}

        \begin{figure}[htbp]
            \begin{center}
                \includegraphics[width=0.45\textwidth]{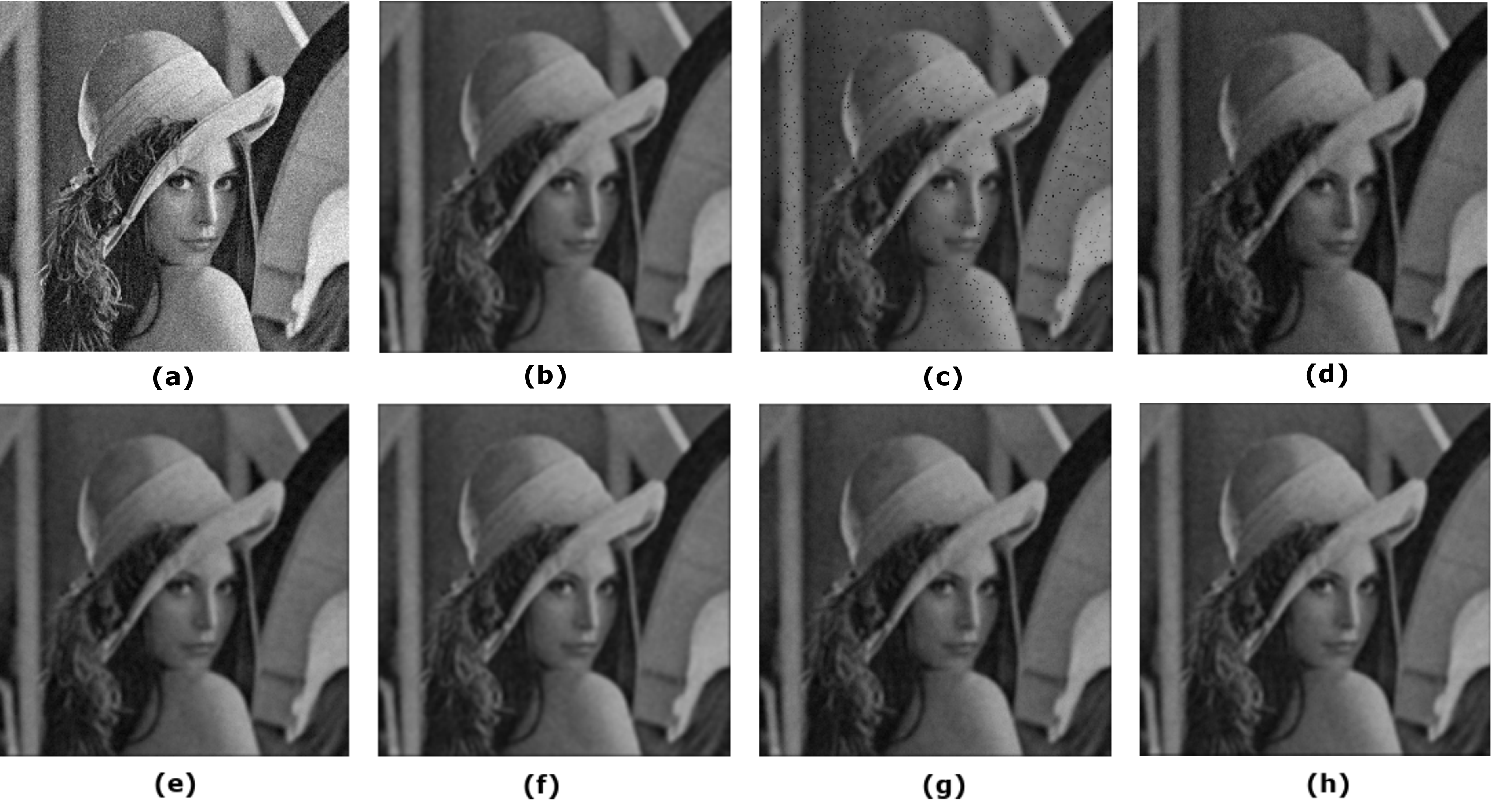}
                \caption{Gaussian image smoothing (a) Original image, (b) Original image with noise, (c) RAP-CLA, PSNR=29.366dB \& SSIM=0.7814 (d) SARA, PSNR=26.79dB \& SSIM=0.787 (e) BCSA, PSNR=33.9dB \& SSIM=0.9142 (f) BCSA with ERU, PSNR=37.837dB \& SSIM=0.9482 (g) CESA, PSNR=32.032dB \& SSIM=0.9007 (h) CESA-PERL, PSNR=36.097dB \& SSIM=0.9302}
                \label{fig:lena}
            \end{center}
        \end{figure}

In this section, we study the application of Gaussian smoothing using CESA and CESA-PERL and compare them to other state-of-the-art approximate adders as shown in Figure~\ref{fig:lena}. For this purpose, we take a 256x256 grayscale image of {\textit{Lena}} and apply Gaussian smoothing to it. We take a $5 \times 5$ filter and apply it to a noisy image of {\textit{Lena}}. The original filter has fractional numbers, but for our application using CESA/CESA-PERL, 
we need fixed-point numbers, hence we round them. 
The addition operation in convolution is approximated and the rest of the arithmetic operations are unchanged.

We compare our adder's accuracy with that of other state-of-the-art approximate adders based on the metrics Peak Signal to Noise Ratio (PSNR) and Structural Similarity (SSIM)~\cite{ssim} Index for 32-bit approximate adders with a block size of 8. The PSNRs and SSIMs of all approximate adders are compared with respect to accurate addition in Gaussian smoothing. The results indicate that CESA has a PSNR of 32.032dB and an SSIM of 0.9007 which is 12.3\% better than that of RAP-CLA and SARA combined. This gain can be attributed to the use of parallel carry estimation done by CESA. The PSNR and SSIM values of BCSA and BCSA with ERU are however better than CESA by 10.4\% and 3.25\% respectively because in BCSA and in BCSA with ERU, an Error Reduction Unit is present which inherently provides better results than our technique with a significant area overhead. Coming to CESA-PERL, Figure~\ref{fig:lena} shows that it outperforms all other approximate adders except BCSA with ERU. 
But compared to PERL, ERU consumes higher on-chip area and produces more accurate results at the expense of latency.
     
\subsection{K-Means Clustering}
        \begin{figure}[htbp]
            \begin{center}
                \includegraphics[width=0.45\textwidth]{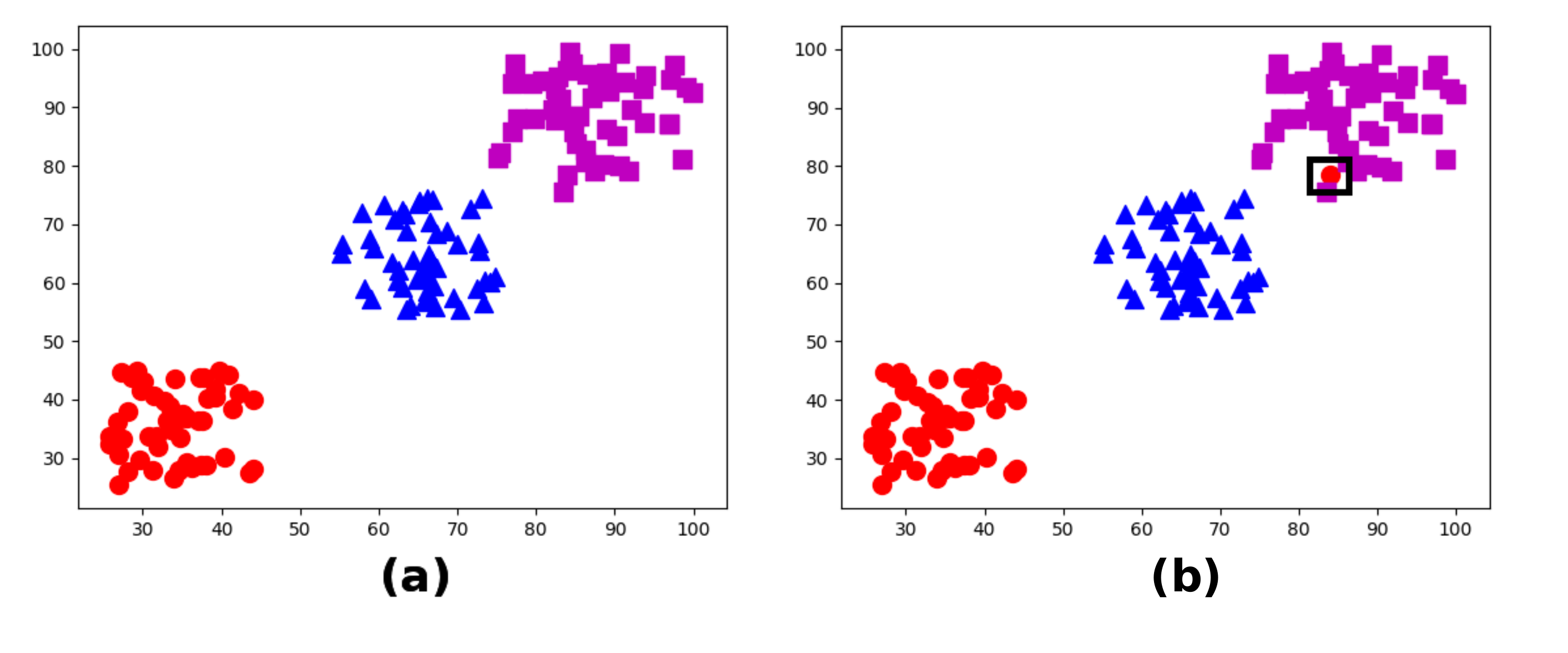}   
                \caption{K-Means (a) CESA-PERL K-Means with (32,8) and (32,16) (b) CESA-PERL K-Means with (32,4). Incorrectly clustered data point is highlighted in black box.}
                \label{fig:kmeas-out}
            \end{center}
        \end{figure}
We have evaluated CESA-PERL on K-Means Clustering Algorithm. We have considered a simple dataset containing 150 data points with 3 clusters. For bit size and block size configurations of (32, 8) and (32, 16), our adder performs accurate clustering on the given dataset (Figure~\ref{fig:kmeas-out}. (a)). However, for CESA-PERL configured with bit size of 32 and block size of 4, our results differ from the accurate result by 0.66 (Figure~\ref{fig:kmeas-out}.  (b)). Evaluating CESA on K-Means show similar results with the same incorrect clustering accuracy.  
        
        

        
    \subsection{Analysis on SPEC CPU2006 Benchmarks}
        Towards the measurement of \textit{performance improvement}, we have used GEM5 statistics for 7 SPEC CPU2006 benchmarks, as mentioned in Section~\ref{subsec:systems}. Our proposed adders, CESA and CESA-PERL, were used to find the speedup obtained for a program 's addition operations, compared to a baseline case of a conventional system with a ripple carry adder. The speedup obtained using CESA-PERL for bit size, block size configurations of (32, 4); (32, 8) and (32, 16) were 2.57x, 2.03x, and 1.50x respectively, compared to the baseline case. CESA-PERL cannot be used in (32,2) configuration due to the minimum bits requirement as mentioned in Section~\ref{sec:su}. So, we have used CESA adder to obtain the results for the (32, 2) configuration. The speedup obtained was 2.83x.
    

\vspace{-1.5ex}
\section{Related Works}

Approximate adders, in general, are developed for catering to compute-intensive applications that require fast computations. An approximate extension to carry-lookahead adder was proposed through RAP-CLA~\cite{rapcla} which reduced the area of the actual carry-lookahead adder. RAP-CLA, however, suffers delay that is higher than CLA and produces results that are on an average 63.7\% more error-prone than an accurate adder.


The approximate binary adders~\cite{eta, etai} split the input operands into two segments. Here, the LSBs are approximately computed and MSBs are accurately computed, thus producing the result in lesser time. Few other types of approximate adders are based on carry-selection~\cite{high, recon, newapp}. In these adders, every block computes sum assuming carry input equals 0 and 1 similar to conventional carry select adder and one of them is selected based on predicted carry rather than accurate carry. The approximate adders RAP-CLA~\cite{rapcla} and SARA~\cite{sara} speculate carry and 
compute correct carry, thus leading to a costlier design than an accurate adder. In some other approximate adders, the input operands are divided into various segments. In~\cite{etaii,bcsa}, the sum of each segment is computed independently with blocks being computed with either accurate or approximate methods.

It is worth noting that \cite{sara, rapcla} employ multiplexers in their design which leads to higher consumption of on-chip area. Hence we investigate two designs, CESA and CESA-PERL, CESA doesn't require multiplexing but has lower accuracy for larger computations whereas CESA-PERL with the addition of multiplexers provides higher accuracy for both small and large computations. Our solution for the fundamental adder architecture (CESA) differs from the existing solutions from the point that we do not use any multiplexer in it leading to lower requirement of space with better accuracy. This design while providing good results on smaller block sizes generates some error on larger computations. To counter that, we also propose a rectification logic (PERL) which would negate the effect of propagating error to a great extent at the cost of an additional area overhead. In this case, we make use of one multiplexer per block for CESA-PERL to select the carry input.

\section{Conclusion and Future Work}
In this paper, we propose an approximate carry estimating simultaneous adder called CESA. It is based on a nearly accurate carry estimation of carry-out using a carry estimator circuit. It has significantly lower power consumption, delay and area overhead than other state-of-the-art approximate adders. Moreover, we also propose a rectification logic PERL which yields more accurate results for larger computations. In the near future, we plan on extending this work to incorporate hardware support for the addition of signed integers alongside floating point numbers and implement a subsequent compiler design to use the same. 

\bibliographystyle{unsrt}  
\bibliography{bib/bibtex/template}  

\begin{thebibliography}{10}

\bibitem{sara}
W.~{Xu}, S.~S. {Sapatnekar}, and J.~{Hu}.
\newblock {A Simple Yet Efficient Accuracy-Configurable Adder Design}.
\newblock {\em IEEE Transactions on VLSI Systems}, pages 1112--1125, 2018.

\bibitem{bcsa}
F.~{Ebrahimi-Azandaryani}, O.~{Akbari}, M.~{Kamal}, A.~{Afzali-Kusha}, and
  M.~{Pedram}.
\newblock Block-based carry speculative approximate adder for energy-efficient
  applications.
\newblock {\em IEEE Transactions on Circuits and Systems II: Express Briefs},
  67(1):137--141, 2020.

\bibitem{approx}
J.~{Han} and M.~{Orshansky}.
\newblock Approximate computing: An emerging paradigm for energy-efficient
  design.
\newblock In {\em 2013 18th IEEE European Test Symposium (ETS)}, pages 1--6,
  2013.

\bibitem{ansari}
M.~S. {Ansari}, B.~F. {Cockburn}, and J.~{Han}.
\newblock A hardware-efficient logarithmic multiplier with improved accuracy.
\newblock In {\em 2019 Design, Automation Test in Europe Conference Exhibition
  (DATE)}, pages 928--931, 2019.

\bibitem{sys1}
A.~Raha and V.~Raghunathan.
\newblock {Towards Full-System Energy-Accuracy Tradeoffs: A Case Study of An
  Approximate Smart Camera System}.
\newblock In {\em Proceedings of the 54th Annual Design Automation Conference},
  DAC 17, pages 1--6, 2017.

\bibitem{soft2}
M.~Samadi, D.A. Jamshidi, J.~Lee, and S.~Mahlke.
\newblock {Paraprox: Pattern-based Approximation for Data Parallel
  Applications}.
\newblock In {\em 19th ACM Architectural Support for Programming Languages and
  Operating Systems (ASPLOS)}, page 35–50, 2014.

\bibitem{metric}
J.~{Liang}, J.~{Han}, and F.~{Lombardi}.
\newblock New metrics for the reliability of approximate and probabilistic
  adders.
\newblock {\em IEEE Transactions on Computers}, 62(9):1760--1771, 2013.

\bibitem{rapcla}
O.~{Akbari}, M.~{Kamal}, A.~{Afzali-Kusha}, and M.~{Pedram}.
\newblock {RAP-CLA: A Reconfigurable Approximate Carry Look-Ahead Adder}.
\newblock {\em IEEE Transactions on Circuits and Systems II: Express Briefs},
  pages 1089--1093, 2018.

\bibitem{gem5}
N.~Binkert, B.~Beckmann, G.~Black, S.K. Reinhardt, A.~Saidi, A.~Basu,
  J.~Hestness, D.~R. Hower, T.~Krishna, S.~Sardashti, R.~Sen, K.~Sewell,
  M.~Shoaib, N.~Vaish, M.D. Hill, and D.A. Wood.
\newblock {The Gem5 Simulator}.
\newblock {\em SIGARCH Comput. Archit. News}, page 1–7, Aug 2011.

\bibitem{ssim}
Z.~{Wang}, A.~C. {Bovik}, H.~R. {Sheikh}, and E.~P. {Simoncelli}.
\newblock Image quality assessment: from error visibility to structural
  similarity.
\newblock {\em IEEE Transactions on Image Processing}, 13(4):600--612, 2004.

\bibitem{eta}
N.~{Zhu}, W.~L. {Goh}, W.~{Zhang}, K.~S. {Yeo}, and Z.~H. {Kong}.
\newblock {Design of Low-Power High-Speed Truncation-Error-Tolerant Adder and
  Its Application in Digital Signal Processing}.
\newblock {\em IEEE Transactions on VLSI Systems}, pages 1225--1229, 2010.

\bibitem{etai}
N.{ Zhu}, W.~L. {Goh}, and K.~S. {Yeo}.
\newblock {An enhanced low-power high-speed Adder For Error-Tolerant
  application}.
\newblock In {\em {Proceedings of the 12th International Symposium on
  Integrated Circuits}}, pages 69--72, 2009.

\bibitem{high}
K.~{Du}, P.~{Varman}, and K.~{Mohanram}.
\newblock {High performance reliable variable latency carry select addition}.
\newblock In {\em {Design, Automation Test in Europe Conference Exhibition
  (DATE)}}, pages 1257--1262, 2012.

\bibitem{recon}
R.~{Ye}, T.~{Wang}, F.~{Yuan}, R.~{Kumar}, and Q.~{Xu}.
\newblock On reconfiguration-oriented approximate adder design and its
  application.
\newblock In {\em 2013 IEEE/ACM International Conference on Computer-Aided
  Design (ICCAD)}, pages 48--54, 2013.

\bibitem{newapp}
J.~{Hu} and W.~{Qian}.
\newblock A new approximate adder with low relative error and correct sign
  calculation.
\newblock In {\em 2015 Design, Automation Test in Europe Conference Exhibition
  (DATE)}, pages 1449--1454, 2015.

\bibitem{etaii}
N.~{Zhu}, W.~L. {Goh}, and K.~S. {Yeo}.
\newblock {Ultra low-power high-speed flexible Probabilistic Adder for
  Error-Tolerant Applications}.
\newblock In {\em {International SoC Design Conference}}, pages 393--396, 2011.

\end{thebibliography}

\end{document}